\documentclass[a4paper,12pt,oneside]{article}
\usepackage{graphicx}
\usepackage{rotating} 
\usepackage{epsfig} 
\usepackage{amsmath}
\usepackage{amssymb}
\usepackage{euscript}
\usepackage[hyperfootnotes=false]{hyperref}
\usepackage[all]{hypcap}
\usepackage{url}
\begin{document}

\title{
Non-LTE Modelling of the Structure and Spectra of the Hot Accretion Spots on the Surface of Young Stars}

\author{A.V.\,Dodin}

\date{ \it \small
Sternberg Astronomical Institute of Moscow State University,
Universitetskij prospekt 13, Moscow, 119992 Russia
\footnote {E-mail: dodin\_nv@mail.ru}
}

\maketitle

\bigskip


\bigskip

%%%%%%%%%%%%%%%%%%%%%%%%%%%%%%%%%%%%%%%%%%%%%%%%%%%%%%%%%%%%%%%%%%%%%%%%%%%%

\section*{Abstract}
The paper describes the modelling of the structure and spectra of the hot accretion spots on the surface of young stars with taking into account departures from LTE for hydrogen and helium.
It has been found that the existence of the ram pressure of the in-falling gas at the outer boundary of the hot spot leads to the Stark broadening of the hydrogen line profiles up to FWHM$\sim$1000 km\,s$^{-1}$ at the considered accretion parameters. It is shown that taking into account departures from LTE for atoms and ions of carbon and oxygen  does not lead to noticeable changes in the structure of the hot spot.

\section*{Introduction}

Classical T Tauri stars (CTTS) are young (ages $<10^{7}$yr) stars with masses $(M\le 3\,M_\odot)$ at the stage of gravitational contraction to the main sequence whose activity is attributable to the magnetospheric accretion
of matter from a protoplanetary disk. The optical spectrum of CTTS consists of the photospheric spectrum
of a late-type star on which emission lines with nontrivial profiles of variable shape are superimposed.
Generally speaking, the emission lines of CTTS consist of two components, narrow (FWHM $\sim 30$ km\,s$^{-1}$) and  broad (FWHM $> 100$ km\,s$^{-1}$) ones, that are formed in different spatial regions -- see e.g.  Batalha et al.  (1996), Dodin et al. (2012). The flux ratio of these components is different for different lines and changes
with time; for the same line, it changes from star to star.

Scarcely anyone doubts that the magnetospheric accretion theory is able to explain the profiles, intensity and complicated variability of the emission lines in the spectra of CTTS, taking into account a nontrivial kinematics of the gas and a variety of physical conditions in the immediate vicinity of the star, which are predicted by this theory (K{\"o}nigl, 1991; Romanova et al., 2004). In the frame of the magnetospheric model the accreting gas near the star moves along the field lines. However, the gas must first reduce its angular velocity down to the velocity of the stellar magnetic field that must be accompanied by heating the gas and by the outflow formation at the boundary of the magnetosphere. Typical velocities at the magnetospheric boundary $\sim100$ km\,s$^{-1},$ hence this region is well suited for formation of the broad components of emission lines (Gomez de Castro, von Rekowski, 2011).

The gas frozen in the magnetic field slides down toward the star, being accelerated by gravity, and this gas may also manifest itself as broad components (see e.g. Petrov et al., 2014), which can be red-shifted up to the free-fall velocity at the stellar surface:
%   _________________________________________________________________________
%  |                                                                         |
   $$   V_0=\sqrt{\frac{2GM_*}{R_*}}\left(1-\frac{R_*}{R_0}\right)^{1/2},   $$   
%  |_________________________________________________________________________|
%
where $M_*$, $R_*$ are stellar mass and radius , $R_0$ is a distance, from which the gas starts to fall freely along the field lines.

Having reached the dense layers of the stellar atmosphere, matter is decelerated in the accretion shock (AS), converting into the heat the main part of its kinetic energy, the flux of which is equal to
%           _________________________________________________
%          |                                                 |
           $$   F_{acc} =  \frac{\mu m_p N_0 V_0^3}{2},     $$    
%          |_________________________________________________|
%
where $N_0$, $V_0$ are pre-shock gas number density and velocity, $\mu=1.26$ is the average molecular weight of one atomic nucleus for solar elemental abundances, $m_p$ is the proton mass.  The hot post-shock matter cools down gradually radiating its thermal energy in the UV and X-ray spectral bands and settles down to the stellar surface while reducing its velocity practically to zero. This radiation and radiation of free-falling gas irradiate the underlying stellar atmosphere, producing so-called hot spot, which manifests itself in spectra as narrow components of the emission lines, and which leads to reduction of equivalent widths of absorption lines in spectra of CTTS in comparison with spectra of main-sequence stars of the same spectral types.

Radiation of the hot spot in continuum had been calculated by Calvet and Gullbring (1998), then numerous spectral lines were accounted for by Dodin and Lamzin (2012) in LTE calculations of the structure and spectra of the hot spot. As it was argued in subsequent work (Dodin et al., 2013) lines of He\,II showed large departures from LTE and could be useful for diagnostic of the shock. The large departures from LTE in He\,I and He\,II, which are main absorbers of the external radiation, can change the conditions in the atmosphere, however Dodin et al. (2013) have taken the thermal structure and the radiation field from LTE model. Such approximation turned to be insufficient to use He\,I lines for diagnostic purposes that gave impetus to the self-consistent non-LTE calculations.

Modelling of the structure and spectra of the hot spot being considered in this work and in the paper of Dodin 
and Lamzin (2012) is one of the problems about heating of the atmosphere by external radiation, which have been repeatedly solved by various authors (e.g. Mitskevich, Tsymbal, 1992; Sakhibullin, Shimanskii, 1996; G\"unther, Wawrzyn, 2011).
However in the case of CTTS, the problem has a number of features, which eventually qualitatively change the results.
The main feature is a ram pressure on the outer boundary of the atmosphere caused by the in-falling gas that must result in higher density in the irradiated atmosphere and in smaller departures from LTE in comparison with the case when 
the external pressure is absent. Moreover, in our case the source of ionizing radiation is directly adjacent to the atmosphere, irradiating it from all sides simultaneously in comparison with binary systems, where the source of external radiation is located  at a considerable distance from the star, and therefore, the radiation penetrates the atmosphere from a considerably smaller solid angle.

    %%%%%%%%%%%%%%%%%%%%%%%%%%%%%%%%%%%%%%%%%%%%%%%%%%%%%%%%
          %%%%%%%%%%%%%%%%%%%%%%%%%%%%%%%%%%%%%%%%

\section*{Calculation methods}

In this work, the vertical structure of the accretion zone will be conventionally divided into three parts: 
the pre-shock region, the post-shock region and the hot spot. If the pre-shock and post-shock regions are separated by the shock front, then the boundary between the post-shock region and the hot spot cannot be strictly demarcated. Therefore, the post-shock region is defined here as the post-shock region calculated by Lamzin (1998).
The gas in this region is transparent in the continuum and gradually radiates almost all energy produced in the shock, practically reaching the statistical and radiative equilibrium at the bottom boundary of the region, which simultaneously is an upper boundary of the hot spot. The optical depths considered in the paper are calculated beginning from this boundary, and therefore, they do not cover the pre- and post-shock regions. These regions are considered as external envelopes and accounted for separately. A self-consistent solution of all regions is still too complex computational problem, and at this stage the mutual influence of pre-shock gas and hot spot is taken into account by successive approximations in two steps: first, the structure and spectra of the pre-shock region are calculated using a primitive model of the hot spot, then the obtained radiation is used in the modelling of the hot spot itself (see the Appendix in the paper of Dodin and Lamzin, 2012). The next iteration leads to a change in the integral flux of about 5\% in the worst case, and therefore, does not performed bearing in mind the remaining uncertainties of similar scale.

In the frames of such division of the accretion zone, the hot accretion spot will be considered as a plane-parallel stellar atmosphere in hydrostatic and radiative equilibrium.
Thus, we ignore a residual motion of the gas and an advection from the post-shock in the hot spot as well as within the hot spot. To compute the model of such atmosphere, a numerical self-consistent solution of the system of equations described below was carried out.

1. {\it The equation of hydrostatic equilibrium} is solved taking into account the ram pressure of the in-falling gas. The equation considers only the gas and radiation pressure. A pressure, caused by a magnetic field or a turbulence, is ignored.

2. {\it The Boltzmann and Saha equations} for atoms and ions of elements from H to Zn. 

3. {\it The statistical equilibrium equations } for atoms and ions treated in non-LTE. The equations are solved only for a few user-defined elements, the corresponding atomic data are given in the next section.

4. {\it The radiative transfer equation} is solved taking into account the incident radiation at the upper boundary.

To construct a model, which would adequately describe the real stellar atmospheres, all possible opacity sources should be included correctly. Our models were calculated taking into account line (b-b) and continuous opacity (b-f, f-f) for elements treated in LTE and non-LTE, opacity related with $H^-$, the Rayleigh and Thomson scattering. Molecular opacity is not included, hence the program cannot be applied to cool stars, however in the considered case of the hot spot molecules must be absent. 
The line blanketing is included by the opacity sampling method for a number of frequency points more than 60\,000 in the case of calculation of the atmospheric structure. The number of frequency points between $3\,000-12\,000${\AA} was increased up to 180\,000 for the final calculation of spectra that allows to calculate roughly profiles of individual lines. The line profiles of non-LTE-elements were calculated with good accuracy on nonuniform fine grids constructed for each line.

\begin{figure}
\begin{center}
\includegraphics[scale=0.6]{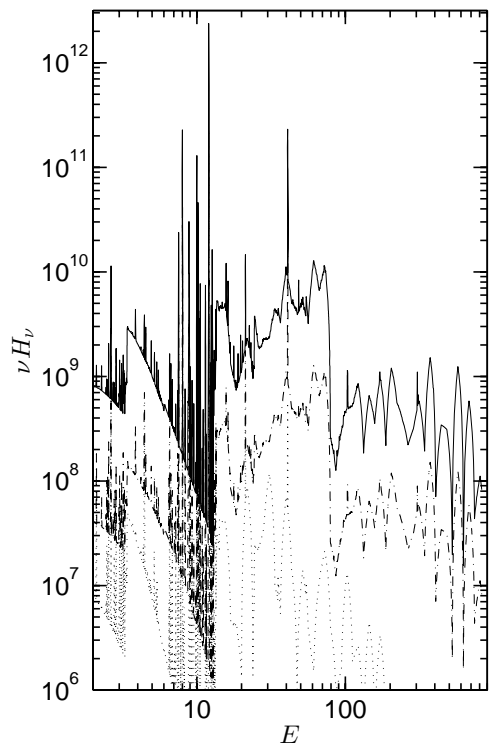}
\end{center}
\caption{Spectra of the external radiation for various accretion parameters.
The dotted curve is for $V_0=200$~km\,s$^{-1},$ $\log N_0 = 11.5;$ 
the dash-dotted curve is for $V_0=400$~km\,s$^{-1},$ $\log N_0 = 11.5;$
the solid curve is for $V_0=400$~km\,s$^{-1},$ $\log N_0 = 12.5.$
The product of frequency and Eddington flux [erg/(s$\times$cm$^2\times$sr)] 
is shown as a function of photon energy E [eV].
}\label{extrad}
\end{figure}

The external radiation is calculated in the similar manner as it was done in the previous work (Dodin and Lamzin, 2012). The only exception is the frequency grid, because now we can choose the most suitable grid. This grid is a combination of grids, on which various components of the external radiation were calculated. There are two such components. The first part is the shock radiation, which has been calculated by Lamzin (1998) in the form of pseudo-lines and signal lines. Since the profiles of signal lines were not calculated in that paper, the doppler shape with a width and redshift of $V_0/8$ was assumed. Test runs showed that the result does not depend on parameters of these profiles. The second component of the external radiation is a radiation emitted by the pre-shock region, which was calculated with the program Cloudy (Ferland et al., 1998) in the same way as it was done in the paper of Dodin and Lamzin (2012). A few examples of the spectra of the external radiation are shown on the Fig.\,\ref{extrad}. Neither data from paper of Lamzin (1998) nor the program Cloudy allows to determine the angular distribution of the intensity, therefore, we will assume, as in the paper of Dodin and Lamzin (2012), that the external radiation is isotropic.

5. {\it The equations of radiative equilibrium and conservation of total (convective and radiative) energy flux.} Convective energy transport is treated using the mixing length theory with $\alpha = 1.25.$ 

The balance between heating and cooling and constant energy flux through the atmosphere in the case of only radiative energy transport are achieved by the temperature correction from the paper of Dreizler (2003). The Avrett-Krook temperature correction modified for convection is used in convection zone (Kurucz, 1970; Castelli, Kurucz, 2004).

The computer program, which solves all these equations, is based on the DETAIL code (Butler, Giddings, 1985),  which allows to solve self-consistently the equations (2 -- 4) for a given atmospheric structure 
by means of accelerated $\Lambda$-iteration.
At the first iteration these equations are solved for an initial model, which is taken from LTE calculations described by Dodin and Lamzin (2012). Then, the model is corrected to satisfy the conditions (1, 5). The corrected model is used to obtain a new solution of the equations (2 -- 4) and to calculate a new corrections. The iterations are repeated until the equations are satisfied to the desired accuracy.
A more detailed description of the program, tests and comparison with other codes can be found on web page\footnote{\href{http://lnfm1.sai.msu.ru/~davnv/hotspot}{\url{lnfm1.sai.msu.ru/~davnv/hotspot}}}.

\section*{Atomic data}

Non-LTE calculations require carefully selected atomic data, the volume of which is greatly increased compared with the calculations in LTE. The atomic data for hydrogen and helium were selected following the recommendations of Przybilla, Butler (2004) and  Przybilla (2005). Testing of the atomic data and a comparison with other data can be found on web page$^1.$ The calculations of the structure and spectra were carried out with one and the same atomic data set. The atomic data set needed to calculate LTE models is equal to the corresponding data set used for the non-LTE calculations. All spectral lines of all elements are calculated assuming complete frequency redistribution.

\subsection*{Hydrogen atom}
The testing showed that a model of hydrogen atom, which takes into account all levels up to $n=20,$ is sufficient to calculate the structure and spectra of various stellar atmospheres, in particular, to calculate the spectrum of the hot spot. The ionization enegry of each level is calculated using the Rydberg formula, taking into account a reducing the ionization energy by the energy of 21st level, at which the continuum is assumed to start:
$$ E_{H}  = R_{H}\left(\frac{1}{n^2}-\frac{1}{21^2}\right). $$
Test calculations of models of stellar atmospheres without an external radiation showed  the advisability of such procedure, however, in the case of considered problem, the models and spectra, calculated with and without the lowering of the ionization energies, practically do not differ from each other (less than 1\% for $H_{\alpha}$ -- $H_{\delta}$). 
Wavelengths of spectral lines are calculated from the levels energies, except the Balmer lines, for which the values from the VALD database (Kupka et al., 1999) were used.

The oscillator strengths are calculated by using the formulae from Berestetskij et al. (1989).
Stark broadening was calculated for line profiles corresponding to the transitions from $n=1-4$ to overlying levels: the tables of Stehle\&Hutcheon (1999) were used for the transitions from $n=1-3$ to $n\leq7$, the theory of Griem (1960) as implemented by Auer\&Mihalas (1972) was applied for the rest lines with a lower level $n=1-4.$ The rest lines are included, assuming Doppler profiles.

Taking into account a big number of lines in the atomic model 
and Stark broadening for lines near series limits is necessary to reproduce 
real shapes of stellar spectra near the ionization threshold. Otherwise, the integral flux can be erroneous that leads to errors in the model of the atmosphere. 

Electron-impact excitation rates are calculated as in the paper of Przybilla, Butler (2004) and correspond to model E from this paper.

Electron-impact ionization rates are evaluated
according to Johnson (1972). 

Photoionization cross sections and the f-f opacity are calculated 
applying hydrogenic expressions (Mihalas, 1978) with Gaunt 
factor $g_{II}$ as in the paper of Przybilla, Butler (2004).

\subsection*{Helium atom}
The atomic model of He\,I includes all states up to principal quantum number $N_{max}=10.$ Energies of levels are adopted from the NIST database and reduced on an energy of the levels with $n=11$ as it was done before, in the case of hydrogen atom. As it was done in the paper of Przybilla (2005), all states up to $n=5,$ which are characterized by $L$ and $S,$ are treated individually, with the remainder grouped over $L$ into combined levels for each $n$ in the singlet and triplet spin systems. Test runs showed that a detailed consideration of the atomic levels up to $n=7$ did not lead to significant changes in the spectral lines, which are commonly used for comparisons with observations.

Wavelengths and oscillator strengths of spectral lines are adopted from the NIST database and supplemented by data from the NORAD\footnote{\href{http://www.astronomy.ohio-state.edu/~nahar/}{\url{http://www.astronomy.ohio-state.edu/~nahar/}}} database (Nahar, 2010). Stark broadening is accounted for using the tables of Dimitrijevic, Sahal-Brechot (1984). Lines, which have not been included in the tables, are calculated, assuming Doppler profiles. Many helium lines have a multicomponent structure, which leads to a significant broadening and distortion of the profiles. Hence the profile $\varphi(\lambda)$ is calculated as a weighted mean over all components of the thin structure:
$$\varphi(\lambda) = \frac{\sum{g_if_i\phi(\lambda-\lambda_i)}}{\sum{g_if_i}},$$
where $\phi$  is the profile of $i$-th component with a wavelength $\lambda_i$ and $gf$-value $g_if_i.$

Electron-impact excitation rates for all transitions from $n=1,2$ (except $2p\,^1P$) to $n=2-5$ are adopted from Bray et al. (2000). Additional transitions are treated according to Mihalas \& Stone (1968), and for the remainder of the optically forbidden transitions, the semiempirical Allen formula (Allen, 1973) is applied.

Collisioanl ionization is accounted for according to Mihalas, Stone (1968).

Photoionization cross-sections for all levels with $n\le7$ are adopted from the NORAD database (Nahar, 2010).
Photoionization cross sections for the levels with $n=8-10$ are evaluated 
applying hydrogenic expressions.

\subsection*{Ionized helium atom}

The model of He\,II atom includes all states up to principal quantum number $n=40.$
The energy for each state is defined in the similar way as for hydrogen. 
Oscillator strengths are hydrogenic. Stark broadening was calculated for line profiles corresponding to the transitions from $n=1-4$ to overlying levels using the theory of Griem (1960) as implemented by Auer and Mihalas (1972). The profiles for the lines 4686\AA{} and 1640\AA{} are calculated using the tables of Sch{\"o}ning, Butler (1989) taking into account the multicomponent structure of the lines. 
The rest lines are included, assuming Doppler profiles.

Electron-impact excitation rates for all transitions from $n=1,2$ to $n=2-5$ are adopted from the CHIANTIv5.0 database (Dere et al., 1997, Landi et al., 2006).  The theory of Percival and Richards (1978) is applied for the transitions between levels with $n\ge5.$  A scaled fit to Sampson \& Golden (Mihalas, 1972) is used for the rest transitions. 

Electron-impact ionization of the levels with $n\le7$ is accounted for
according to Clark et al. (1991), and for the remainder levels, 
the Seaton (1962) approximation is applied.

Photoionization cross-sections for He\,II are calculated applying hydrogenic expressions.

\subsection*{The atomic models for C\,I -- IV, O\,I -- IV}

To qualitatively evaluate the impact of non-LTE effects in carbon and oxygen on the atmospheric model,  simplified models of atoms C\,I -- IV, O\,I -- IV were used. The energy levels used in the calculations
are given in the Tables \ref{ctable} and \ref{otable}. In the case of calculations of the level populations, close energy states are grouped into combined levels, but the internal structure of these combined levels is taken into account at calculations of the spectrum in the similar way as it was done for the fine structure of helium lines. Such procedure allows to take into account a big number of spectral lines, using a small number of atomic levels.

Energies and statistical weights of all states are adopted from the NIST database. Wavelengths, oscillator strengths and damping constants are adopted from the VALD database (Kupka et al., 1999) and supplemented by data from the NIST database and, in the case of C\,II, from the paper of  Fischer, Tachiev (2004).
 Electron-impact excitation rates
for C\,I  are taken from the paper of Reid (1994); 
for C\,II --  Wilson et al. (2005); 
for C\,III --  Mitnik et al.(2003); 
for C\,IV --   Griffin et al.(2000); 
for O\,I -- Barclem (2007), Bhatia, Kastner (1995); 
for O\,II --  Kisielius et al. (2009), McLaughlin, Bell (1994);
for O\,III -- Aggarval, Keenan (1999); 
for O\,IV --  Aggarval, Keenan (2008).            
Collisional excitation for transitions without detailed data is treated using the van Regemorter (1962) approximation in the case of optically allowed transitions, and applying the semi-empirical Allen (1973) formulae in the optically forbidden case.
Photoionization cross-sections for all atoms and ions are adopted from the NORAD database -- see
Nahar, Pradhan (1991, 1994, 1997); Nahar (1994, 1995). The analysis of coefficients for the electron-impact ionization carried out by Avrett and Loeser (2008) showed a significant scatter in the coefficients from different authors that allowed us to use the approximate formula given by Seaton (1962) with the same uncertainty. The charge exchange between oxygen and hydrogen is accounted for according to Arnaud, Rothenflug (1985).

			%%%%%%%%%%%%%%%%%%%%%%%%%%%%

\begin{table}
  \caption{The energy levels of C\,I -- IV }
 \label{ctable}
\begin{center}
\small
\begin{tabular}{|l   r|l   r|l   r|l   r|l   r|}
\hline
$i$&$E$ (cm$^{-1}$)&$i$&$E$ (cm$^{-1}$)&$i$&$E$ (cm$^{-1}$)&$i$&$E$ (cm$^{-1}$)&$i$&$E$ (cm$^{-1}$)\\
\hline
\multicolumn{2}{|c|}{C\,I}& 10(20) & 77679.8  & 10     & 150461.6 & 4  & 137425.7 & 16   & 311721.5 \\
1    &     0.0 &        &  ...      &       & 150466.7 &    & 137454.4 & 17   & 317794.3 \\
     &   16.40 &        & 80834.6  & 11     & 157234.1 &    & 137502.0 &      & 317796.5 \\
     &   43.40 & \multicolumn{2}{|c|}{C\,II} & 12     & 162517.9 & 5  & 145876.1 &      & 317801.3 \\
2    & 10192.6 & 1      &    0.0  &        & 162524.6 & 6  & 182519.9 & 18   & 319720.4 \\
3    & 21648.0 &        &   63.42  & 13     & 166967.1 & 7  & 238213.0 & 19   & 321411.3 \\
4    & 33735.2 & 2      & 43003.3  &        & 166990.7 & 8  & 247170.3 &      & 321426.7 \\
5    & 60333.4 &        & 43025.3  &        & 167035.7 & 9  & 258931.3 &      & 321450.1 \\
     & 60352.6 &        & 43053.6  & 14     & 168123.7 & 10 & 259705.6 & 20   & 322003.7 \\
     & 60393.1 & 3      & 74930.1  &        & 168124.5 &    & 259711.2 &      & 322009.6 \\
6    & 61981.8 &        & 74932.6  & 15     & 168729.5 &    & 259724.3 &      & 322018.0 \\
7    & 64086.9 & 4      & 96493.7  &        & 168748.3 & 11 & 270010.8 & \multicolumn{2}{|c|}{C\,IV}\\
     & 64089.9 & 5      & 110624.2 & 16     & 168978.3 &    & 270011.9 & 1    &     0.0   \\
     & 64091.0 &        & 110665.6 &        & 168978.3 &    & 270014.7 & 2    & 64484.0  \\
8(10)& 68856.3 & 6      & 116537.7 & \multicolumn{2}{|c|}{C\,III}& 12 & 276482.9 &      & 64591.7  \\
     &    ...  & 7      & 131724.4 &  1     &     0.0 & 13 & 308216.6 & 3    & 302849.0 \\
     & 73975.9 &        & 131735.5 &  2     & 52367.06 &    & 308248.9 & 4    & 320050.1 \\
9    & 75254.0 & 8      & 142027.1 &        & 52390.75 &    & 308317.3 &      & 320081.7 \\
     & 75255.3 & 9      & 145549.3 &        & 52447.11 & 14 & 309457.2 & 5    & 324879.8 \\
     & 75256.1 &        & 145550.7 &  3     & 102352.0 & 15 & 310006.3 &      & 324890.3 \\
\hline
\multicolumn{10}{p{14cm}}
 {\footnotesize {\bf Note. } $i$ is an ordinal number of combined level, $E$ is the energy of atomic states grouped in the level $i$. The number in parentheses is the number of the grouped states.}
\end{tabular}
\end{center}
\end{table}

\begin{table}
  \caption{The energy levels of O\,I -- IV }
 \label{otable}
\begin{center}
\small
\begin{tabular}{|l   r|l   r|l   r|l   r|l   r|}
\hline
$i$&$E$ (cm$^{-1}$)&$i$&$E$ (cm$^{-1}$)&$i$&$E$ (cm$^{-1}$)&$i$&$E$ (cm$^{-1}$)&$i$&$E$ (cm$^{-1}$)\\
\hline
\multicolumn{2}{|c|}{O\,I}&\multicolumn{2}{|c|}{O\,II}&\multicolumn{2}{|c|}{O\,III}& &  &\multicolumn{2}{|c|}{O\,IV}\\
1    &    0.0  & 1    &     0.0  & 1  &   0.0   & 11 & 273081.3 & 1 & 0.0 \\
     & 158.265  & 2    & 26810.55  &    &  113.178  & 12 & 283759.7 &   & 385.9 \\
     & 226.977  &      & 26830.57  &    &  306.174  &    & 283977.4 & 2 & 71439.8 \\
2    & 15867.9  & 3    & 40468.01  & 2  & 20273.27  &    & 284071.9 &   & 71570.1 \\
3    & 33792.6  &      & 40470.00  & 3  & 43185.74  & 13 & 290958.3 &   & 71755.5 \\
4    & 73768.2  & 4    & 119837.2  & 4  & 60324.79  & 14 & 293866.5 & 3 & 126936.3 \\
5    & 76795.0  &      & 120000.4  & 5  & 120025.2  &    & 294002.9 &   & 126950.2 \\
6    & 86625.8  &      & 120082.9  &    & 120053.4  &    & 294223.1 & 4 & 164366.4 \\
     & 86627.8  & 5    & 165988.5  &    & 120058.2  & 15 & 324464.9 & 5 & 180480.8 \\
     & 86631.5  &      & 165996.5  & 6  & 142381.0  &    & 324660.8 &   & 180724.2 \\
7    & 88631.2  & 6    & 185235.3  &    & 142381.8  &    & 324839.0 & 6 & 231537.5 \\
     & 88630.6  &      & 185340.6  &    & 142393.5  & 16 & 324735.7 & 7 & 255155.9 \\
     & 88631.3  &      & 185499.1  & 7  & 187054.0  & 17 & 327229.3 &   & 255184.9 \\
8    & 95476.7  & 7    & 188888.5  & 8  & 197087.7  &    & 327278.3 & 8 & 289015.4 \\
9    & 96225.1  &      & 189068.5  & 9  & 210461.8  &    & 327352.2 &   & 289023.5 \\
10(8)& 97420.6  & 8    & 195710.5  & 10 & 267258.7  & 18 & 329469.8 & 9 & 357614.3 \\
     & ...      & 9    & 203942.3  &    & 267377.1  &    & 329583.9 & 10& 390161.2 \\
     & 97488.5  & 10(6)& 206730.8  &    & 267634.0  &    & 329645.1 &   & 390248.0 \\
     &          &      &   ...     &    &           & 19 & 331821.4 & 11& 419533.9 \\
     &          &      & 206972.7  &    &           &    &           &   & 419550.6 \\
\hline
\multicolumn{10}{p{14cm}}
 {\footnotesize {\bf Note. } $i$ is an ordinal number of combined level, $E$ is the energy of atomic states grouped in the level $i$. The number in parentheses is the number of the grouped states.}
\end{tabular}
\end{center}
\end{table}

%
			%%%%%%%%%%%%%%%%%%%%%%%%%%%%

\subsection*{Atomic data for LTE-elements}

Opacity in numerous lines of atoms and ions from C to Zn is considered in LTE with solar elemental abundances\footnote{The C and O elements can be calculated in LTE as well as in non-LTE, depending on the task.}. To reduce the CPU time, the ions, the concentration of which is less than $10^{-10}$ of the total concentration of all atoms and ions, are ignored.

Energies and statistical weights of the atomic states, which are necessary for the Saha equations, are adopted from the NIST database. To simplify the calculations, the close states were grouped into the combined levels, however the number of the atomic states was  big enough that the obtained ionization degrees of ions with the simplified atomic models differ $\lesssim1\%$ from the corresponding ionization degrees, calculated for atomic models, which include all atomic states given in the NIST database.

All atomic parameters necessary to compute the lines are read from the special files, which are used by the ATLAS12 program (Castelli, 2005) to take into account the line blanketing, and which contain the data approximately for 42 million lines.

Photoionization cross-sections from ground and excited states are adopted from the Opacity Project (TOPBase) for the following elements and ions: N\,I -- V, Ne\,I -- VI, Mg\,I -- VI, Al\,I -- VII, Si\,I -- VI, S\,I -- VII, Ar\,I -- VIII. Photoionization cross-sections for C\,I -- IV, O\,I -- VI are the same as used in the non-LTE calculations.

\section*{The sensitivity of the results to uncertainties in the atomic data and to the angular distribution of the external radiation}

Using insufficiently accurate atomic data in the non-LTE modelling can cause systematic errors. In the case of our non-LTE-elements, the atomic data related with radiative processes can be considered as quite accurate, while the coefficients related with collisional transitions were calculated in some cases using very approximate expressions. Therefore it is necessary to find out the scale of errors that can be made in the model or spectrum of the hot spot due to poor quality of the atomic data.

\begin{figure}
\begin{center}
\includegraphics[scale=0.6]{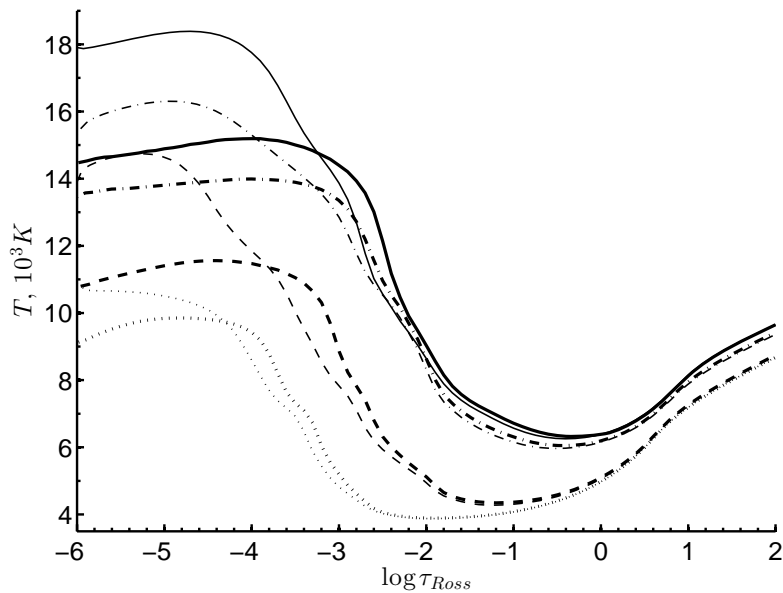}
\end{center}
\caption{Temperature as a function of Rosseland optical depth $\tau_{Ross}$ for LTE (the thin curves) and non-LTE (the thick curves) models of the hot spot at various parameters.
The dotted curves are for $V_0=250$ km\,s$^{-1},$ $\log N_0 = 11.5$; 
the dash-dotted curves are for $V_0=250$ km\,s$^{-1},$ $\log N_0 = 13$;
the dashed curves are for $V_0=400$ km\,s$^{-1},$ $\log N_0 = 11.5$;
the solid curves are for $V_0=400$ km\,s$^{-1},$ $\log N_0 = 12.5$.
The parameters of the underlying star are $T_{eff}=4500$~K, $\log g = 4.0$.}\label{LTE-nLTE}
\end{figure}

Only non-LTE models depend on the uncertainty in the collisional rates, therefore, first of all, let's consider the differences between LTE and non-LTE models that give the scale and character of the changes.
These changes are shown for four models on Fig.\,\ref{LTE-nLTE}. It can be seen from the figure that in the non-LTE case the upper layers of the hot spot are cooler, while the internal layers are hotter in comparison with the LTE-models. Despite the fact that the differences in the dependences of $T(\tau_{Ross})$ between LTE and non-LTE models are $\lesssim20$\%, the spectra differ not only quantitatively but also qualitatively.

The comparison shows that non-LTE effects are least of all manifested in hydrogen lines, namely, the central intensity of the lines in the non-LTE case is reduced by several tens of percent for models with a large $F_{acc}$ and by about two times for models with a low $F_{acc}.$ The line profiles change shape, for instance, the profile of $H_\alpha$ line in the non-LTE models becomes broader in comparison with the LTE case and shows a small absorption feature at the line center (see Fig.\,\ref{profiles}). The differences of spectra between LTE and non-LTE models disappear for high members of the Balmer series.

The principal differences between LTE and non-LTE spectra are observed in He\,I lines and especially in He\,II lines, which are not formed in LTE in the conditions of the hot spot. Such qualitative changes of the spectrum are associated with the presence of the ionizing external radiation in the atmosphere.

\begin{figure}
\begin{center}
\includegraphics[scale=0.6]{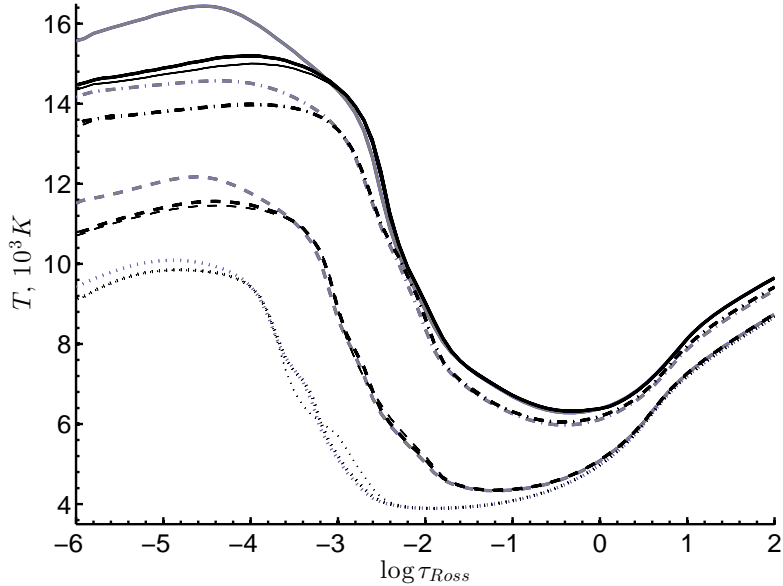}
\end{center}
\caption{Temperature as a function of Rosseland optical depth $\tau_{Ross}$ in the models of hot spots, in which H and He are treated in non-LTE (the thick curves) and in which H, He, C and O are treated in non-LTE (the thin curves). The grey curves are for the models, calculated in non-LTE for H and He and with the angular distribution of the external radiation $I_e(\mu)\propto\mu^{-1}$.  
The parameters of the models are the same as in Fig.\,\ref{LTE-nLTE}.
}\label{LTE-nLTECO}
\end{figure}

Our modelling shows that the external radiation with $\lambda<912$\,{\AA} is primarily absorbed by atoms of H, He\,I, He\,II and only at wavelengths of $230-400$\,{\AA} other elements can be important, primarily the ion C\,III. Hence the models, taking into account departures from LTE for C\,I -- V as well as for hydrogen and helium, were calculated. However, a significant impact of non-LTE effects in carbon on the model or spectrum of hydrogen and helium has not been found. The inclusion of departures from LTE for O\,I -- V leads to similar changes, therefore further addition of less important elements to the non-LTE treatment at the calculation of the atmospheric structure seems like an unjustified complication bearing in mind other uncertainties. Accounting for ions of higher stage of ionization would also be redundant, because the ions with charge $+4$ already have a negligible concentration and the dominant stages of ionization are C\,I -- III and O\,I -- III, depending on depth. Temperature as a function of Rosseland optical depth $\tau_{Ross}$ are shown on the Fig.\,\ref{LTE-nLTECO} for the models, in which H, He, C, O are treated in non-LTE.

Let's consider now how the inaccurate atomic data can impact the spectrum of the hot spot.
Using the electron-impact ionization rates calculated according to Johnson (1972) and by more approximate formula given by Seaton (1962) leads to the spectra, which differ from each other in the Balmer and Pashen lines by $\sim0.1\%.$ Similar calculations for atoms of He\,I -- He\,II lead to the differences in optical lines of He\,I about $0.5\%$ and less than 0.1\% for He\,II. Thus, it can be concluded that 
uncertainties in the electron-impact ionization rates for H\,I, He\,I and He\,II do not impact the structure and spectra of the hot spot.

Electron-impact excitation rates for He\,II for the transition from the levels with $n\ge3$ are calculated by using a scaled fit to Sampson\&Golden (Mihalas, 1972), however the rate coefficients for the transition {"3-4"} can be found in the book of Sobel'man et al. (2002). The comparison shows that the line 4686{\AA}, calculated with Sobel'man's data, is enhanced about two times, while the line 1640{\AA} is weakened by 30\%.
Such uncertainties should be borne in mind when comparing calculations with observations.

The emission of the accretion shock has been calculated by Lamzin (1998) without 
any information about the angular distribution of the intensity, which is necessary 
for our calculations. This introduces an additional uncertainty in the model. 
The true angular distribution of the incident radiation $I_e(\mu)$ must depend 
on the geometry of the accretion column and can be different at different wavelengths, 
nevertheless, on the basis of general considerations, it must be enclosed between 
the isotropic distribution $I_e(\mu)=const$ for $\mu<0$ and the distribution $I_e(\mu)\propto\mu^{-1}$ 
for $\mu<0$ (in both cases $I_e(\mu)=0$ for $\mu\geq0,$ by definition of $I_e$). This problem has been already discussed by Dodin and Lamzin (2012), 
however in the non-LTE case it should be reconsidered, since in case of using
$I_e(\mu)\propto\mu^{-1}$ instead of the isotropic approximation the corresponding 
increasing of the mean intensity in upper layers can change (enhance) non-LTE effects. 
The calculations show that, as in the LTE models (Dodin, Lamzin, 2012), the use 
of $I_e(\mu)\propto\mu^{-1}$ instead of $I_e(\mu)=const$  leads to heating of 
the upper layers by $\Delta T \sim 100$~K for the models with small $F_{acc}$ and 
by $\Delta T \sim 1000$~K in the case of large $F_{acc}$ (see Fig.\,\ref{LTE-nLTECO}). 
While the temperature of more deep layers slightly decreases. The distribution 
$I_e(\mu)\propto\mu^{-1}$ in comparison with $I_e(\mu)=const$ leads to increase 
in intensities of hydrogen and helium lines by $<20\%$ and to decrease in 
the continuous emission by a few percents. The greatest enhancement is observed 
for $H_{\alpha},$ but the differences gradually disappear with increasing a number 
of line in the series.

\section*{The dependence of shape of the spectrum on the accretion parameters}

\begin{sidewaysfigure} 
%\begin{center}
%\includegraphics[scale=0.6]{figures/profiles.eps}
\centerline{\epsfig{file=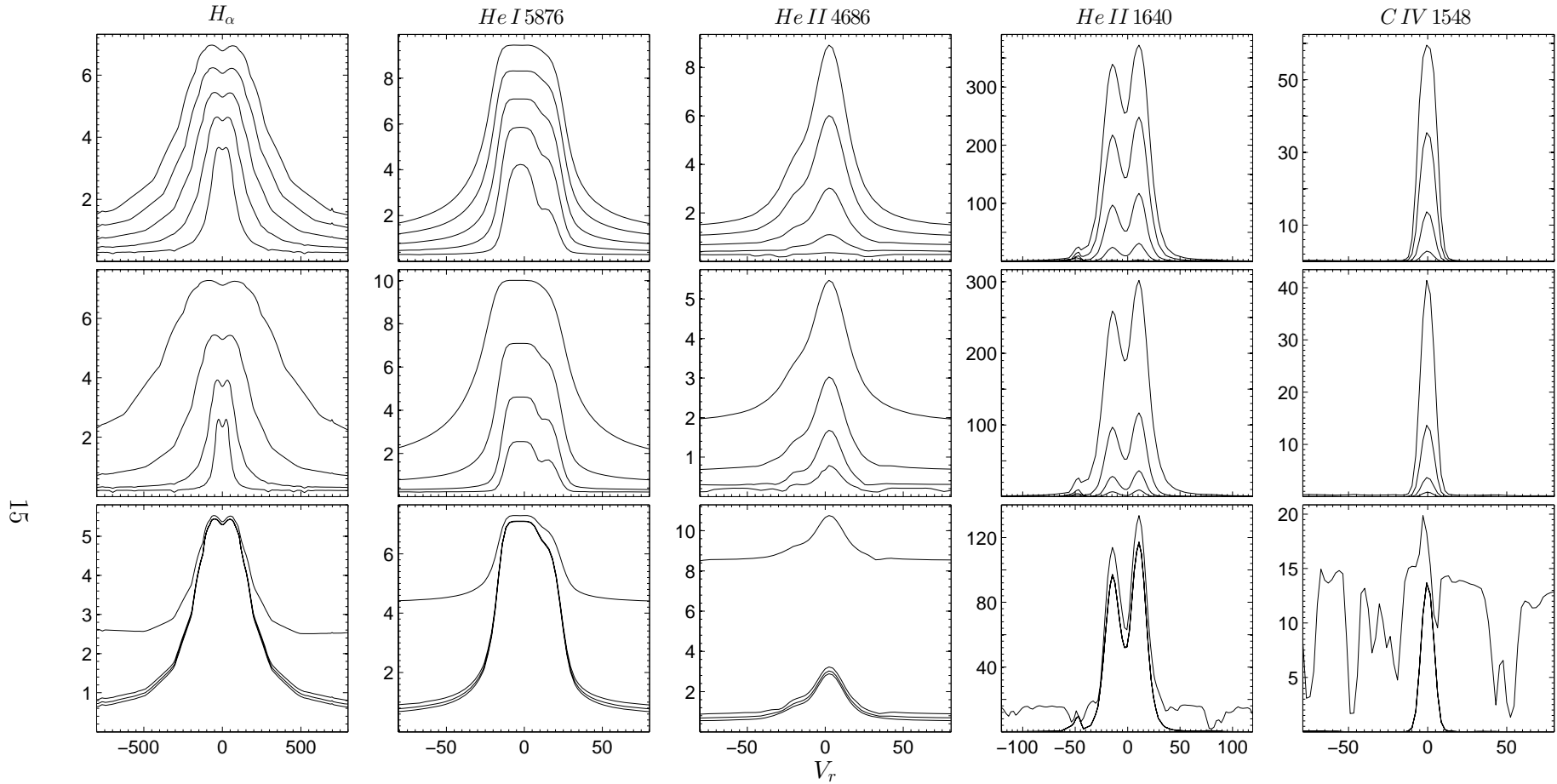, scale=0.68}} 
%\end{center}
\caption{Changes in the line profiles at variation of different parameters.
The vertical axis shows the Eddington flux $H_{\lambda}$ in the units of
$10^{6}$ erg/(s$\times$cm$^2$$\times$\AA$\times$sr). 
The horizontal axis is the radial velocity with respect to the line center in km\,s$^{-1}.$ 
The upper row: $\log N_0=12.5,$ $T_{eff}=4500$~K, $V_0=$200, 250, 300, 350, 400~km\,s$^{-1}.$ 
The middle row: $V_0=300$~km\,s$^{-1}$, $T_{eff}=4500$~K, $\log N_0=$ 11.5, 12.0, 12.5, 13.0. 
The bottom row: $V_0=300$~km\,s$^{-1}$, $\log N_0=$ 12.5, $T_{eff}=$ 4000, 4500, 5000, 10000~K. 
On all panels the flux increases with increase of the varying parameter.}
\label{profiles}
\end{sidewaysfigure}

Having estimated in the previous section the accuracy of the calculated spectra, let's consider the dependence of the spectrum on the basic parameters, which must determine the structure of the hot spot.
The profiles of some important lines are shown on the Fig.\,\ref{profiles} for a typical values of the accretion parameters (Lamzin, 1998) $\log N_0=11.5,12,12.5,13,$ $V_0=200,250,300,350,400$~km\,s$^{-1}$ and for the effective temperatures of the star $T_{eff}=4000,4500,5000,10000$~K. 
Due to limited computational capabilities, only these parameters are varied, while the rest are fixed, namely, the chemical composition is solar, the gravity $\log g = 4.0$, the microturbulence velocity $V_{mic}=2.0$ km\,s$^{-1}.$ 

Changes in the flux of the $H_{\alpha}$ line are shown in the first column on Fig.\,\ref{profiles}.
It can be seen that the typical FWHM of hydrogen lines, being formed in the hot spot, $\gtrsim100$km\,s$^{-1},$ therefore such lines in an observed spectrum will be interpreted as  broad emission components, the formation of which is usually attributed to a circumstellar gas, rather than to a hot spot (see the Introduction). The width of the hydrogen lines is strongly depend on the accretion parameters, increasing with growth of $V_0$ or $N_0.$
It may seem that the profile should be determined by a combination of these variables, for example, only by the ram pressure $P\propto N_0V_0^2$ or only by the total flux of the external radiation $F_{acc}\propto N_0V_0^3,$ however the check shows that it is not. 

As it was expected, the hot spot emits narrow lines of He\,I -- He\,II, which can be 
associated with the narrow emission components in observed spectra of T Tauri stars. 
In comparison with previous work (Dodin et al., 2013), the self-consistent solution 
of the non-LTE problem leads to increase in intensities of He\,I lines up to the 
observed level (see the next section). Note that the center of the He\,I line at 
5876{\AA} shifts to the red when $N_0$ or $V_0$ is increased. This is due to the 
fact that the line consists of several fine-structure components, which are 
sequentially saturated, and therefore, the position of the line center of such 
composite line depends on its optical thickness. On the Fig.\,\ref{profiles} the 
central wavelength of the lines is taken as the $gf$-weighted average wavelength 
of all components of the fine structure, that corresponds to the optically thin case.

The line profiles shown on the Fig.\,\ref{profiles} correspond only to the emission 
of the hot spot, while any emission/absorption in the post- and pre-shock 
gas is not considered.

The pre-shock region manifests itself in the form of broad components, which are shifted in radial velocity by $V_0\cos \alpha,$ where $\alpha$ is an angle between the line of sight and the streamline of the in-falling gas. The pre-shock gas may manifest itself not only in lines, but also in continuum, especially at wavelengths shorter than the Balmer jump, where its contribution to the total flux, according to the calculations with the program Cloudy, can be comparable to the contribution from the hot spot. Hence the calculated spectrum of the hot spot cannot be compared with the observations without taking into account the emission and absorption in the in-falling gas, especially in UV.

The post-shock region produces lines with a small redshift. These lines can contribute to the narrow emission component from the red side of the profile. Due to the high ionization degree this region does not emit the lines of H\,I and He\,I, however it can contribute to He\,II lines at $V_0<300$ km\,s$^{-1},$ when these lines are poorly formed in the hot spot. If the lines of more hot ions are considered, the relative contribution of the post-shock region must be higher.
Our non-LTE calculations of C\,I -- IV show that the CIV 1550 doublet (1548, 1550\AA) formed in the hot spot is significantly weaker than the He\,II 1640 line (see Fig.\,\ref{profiles}), while the observed fluxes in the lines are approximately equal -- see e.g. spectra of CTTS with the explicit narrow component (BP Tau, DS Tau, DN Tau, CY Tau, EG Tau, TWA7, V396 Aur, V397 Aur) in the paper of Ardila et al. (2013). From this it can be concluded that the narrow component of the C\,IV lines is formed not in the hot spot, but rather in the post-shock region, as it was assumed in the paper of Lamzin (2003). 
Two notes should be made here. First, very simplified atomic models were used at the calculation of the C\,IV doublet lines. The test computations show that an increase in the completeness of atomic models for the ions of C\,III -- IV leads to a further weakening of the doublet lines, however a quantitative aspect of the problem requires a more detailed consideration. Second, the uncertainty introduced in the UV spectrum by the pre-shock does not affect this conclusion, because we consider the ratio of two lines, which are close in wavelenths, and therefore, they are equally changed.

The structure of the outer region of the hot spot, where the strong emission lines are formed, is determined mainly by the external radiation and weakly depends on the effective temperature of the underlying star. For this reason, the flux in the emission lines of hydrogen and helium weakly depends on $T_{eff}$ 
(see the bottom row on the Fig.\,\ref{profiles}). 
It means that at fixed accretion rate and parameters $N_0,$ $V_0$ the contribution of the hot spot to the total spectrum decreases with increasing the temperature and radius of the star.
So, in the case of the star with $T_{eff}=10000$~K, $R_*=3R_{\odot},$ at the accretion rate of $10^{-7}$ $M_{\odot}/$yr
the typical equivalent widths of He\,I 5876 and He\,II 4686 lines $\lesssim0.05$\AA, while in the case of the star with $T_{eff}=4500$~K, $R_*=1.5R_{\odot}$ these quantities increase by 30 times.

\section*{Comparison with observations}

As noted in the Introduction, Dodin et al. (2013), using LTE structure of the hot spot, have calculated the non-LTE spectrum of helium, however they failed to simultaneously reproduce the observed equivalent widths of He\,I and He\,II lines. The authors suggested that this problem can be overcome by the self-consistent non-LTE modelling of the structure and spectra of the hot spot that has been done in the present work.

It turns out that the helium problem is able to be solved indeed, however, new difficulties arise, which will be demonstrated below on the example of TW Hya, the spectrum of which was obtained on 2011 May 3 with X-shooter on the 8-m VLT of the European Southern Observatory. The spectrum was retrieved from the ESO archive and reduced with the X-Shooter pipeline (version 2.5.0) following the user manual\footnote{\href{ftp://ftp.eso.org/pub/dfs/pipelines/xshooter/xshoo-pipeline-manual-12.3.pdf}{\url{ftp://ftp.eso.org/pub/dfs/pipelines/xshooter/xshoo-pipeline-manual-12.3.pdf}}}.

To demonstrate the arising problems, we will make two simplifications. 

First, we will assume that the accretion zone
consists of a large number of the hot spots, which are uniformly distributed over the visible hemisphere of the star 
and characterized by the same values of $N_0,$ $V_0$. It allows to use the outward part of $H_{\lambda}$ instead of the intensity $I_{\lambda}(\mu)$ and also to avoid additional free parameters, determining the location and shape of the accretion zone. By the way, Calvet and Gullbring (1998) have used the same approximation. 

Second, we will assume that the pre-shock region is transparent for the radiation of the hot spot, hence its radiation $H_{pre}$ can simply be added to the radiation of the hot spot $H_{hs},$ then the total flux of the system "star+hot spots" can be expressed in the form
$$
F = C\left[f(H_{hs}+H_{pre})+(1-f)H_{*}\right],
$$
where $C$ is a normalization constant, $f$ is a filling factor, $H_*$ is a stellar flux,
calculated with the computer programs ATLAS9 and SYNTHE (Castelli, Kurucz, 2004; Sbordone et al., 2004).
Estimates carried out with the program Cloudy show that at the considered parameters the pre-shock region is indeed transparent in the continuum at least for $\lambda>3650$\,{\AA}.

There is not a reliable technique of determining the fundamental parameters of T Tauri stars, taking adequately into account the accretion and spottedness of these stars. Probably, this is the cause of significant scatter of estimates of $T_{eff},$ which are given by different authors for TW Hya:
$T_{eff}=3400-4200$~K (Vacca, Sandell, 2011; Yang et al., 2005). Hence the models of the star with $T_{eff}=3500-4500$~K and with $\log g = 3.5-4.5$ were calculated for the comparison with the observations,
however the character of the principal contradictions with the observations does not depend on specific parameters and will be illustrated for $T_{eff}=4000$~K and $\log g =4.0$.

\begin{figure}
\begin{center}
\includegraphics[scale=0.7]{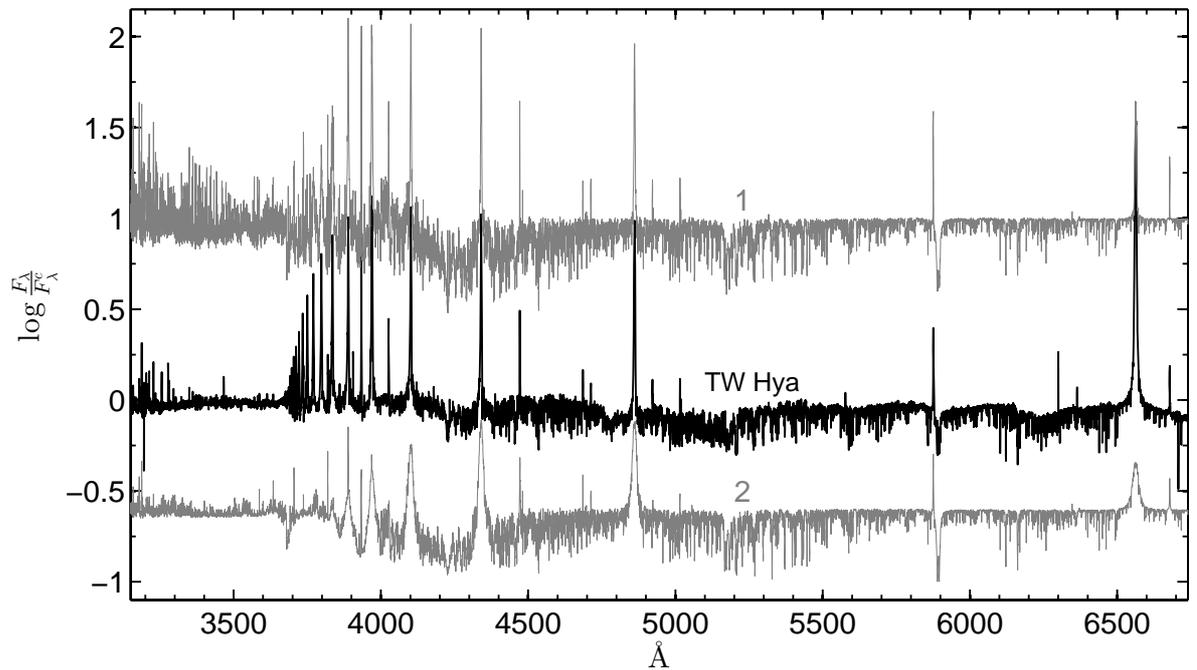}
\end{center}
\caption{The black curve in the center is for logarithm of the observed spectrum of TW Hya (normalized to the continuum). The calculated spectra are shown by the grey curves, which are shifted along $y$ axis for clarity. The spectrum (1) is calculated at the parameters: $T_{eff}=4000$~K, $\log g=4.0$, the microturbulence velocity $V_{mic}=2$~km\,s$^{-1}$, $V_0=300$~km\,s$^{-1}$, $N_0 = 10^{12}$~cm$^{-3}$, $f = 0.11$. 
The spectrum (2) is calculated at the same stellar parameters, but with $V_0=350$~km\,s$^{-1}$, $N_0 = 10^{13}$ cm$^{-3}$, $f = 0.016.$ The theoretical spectra are convolved with a gaussian profile to match widths of absorption lines.}
\label{TWHya}
\end{figure}

\begin{figure}
\begin{center}
\includegraphics[scale=0.55]{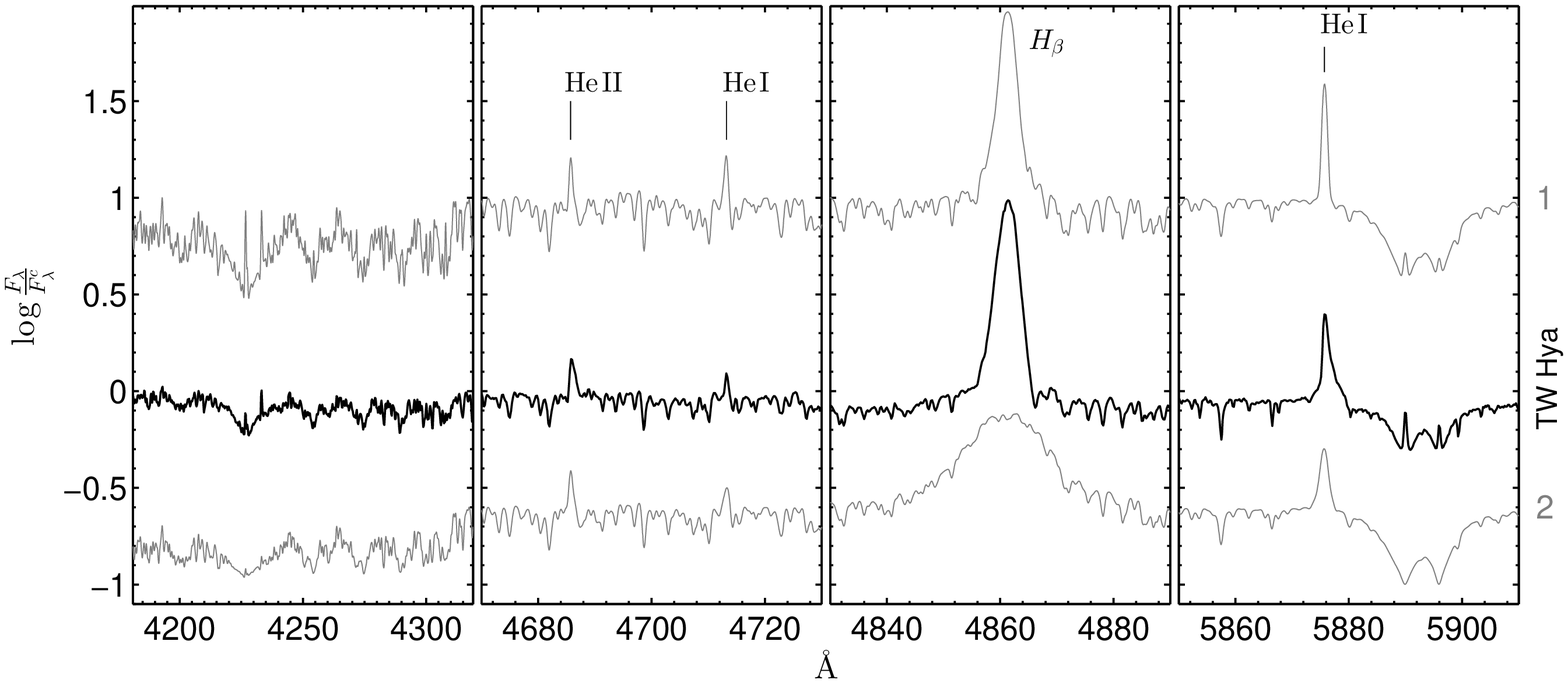}
\end{center}
\caption{Some regions of the spectrum shown in Fig.\,\ref{TWHya}. It can be seen on the first panel that the absorption lines in the spectrum (1) are much stronger than in the observed spectrum.}
\label{TWHya1}
\end{figure}

A comparison of theoretical and observed widths of hydrogen lines 
leads to the conclusion that $N_0\sim10^{12}$ cm$^{-3}.$
At such density the $H_{\beta}$ -- $H_{\delta}$ lines as well as 
the lines of He\,I and He\,II can be reproduced (see the upper curve 
on the Fig.\,\ref{TWHya} and \ref{TWHya1}), while the calculated 
absorption lines turn to be deeper than the observed ones, especially 
in the blue region (see the Fig.\,\ref{TWHya1}). It cannot be excluded that
the discrepancies in line depths are due to our simplifying assumptions, however,
regardless of this, emission lines of hydrogen up to $H_{20}$ are visible in the spectrum of TW Hya, while in the spectra of the hot spot the lines higher than $H_{10}$ are absent at any values of $N_0,$ $V_0.$ 
Moreover, the observed equivalent width of the $H_{\alpha}$ emission line in the spectrum of TW Hya is considerably larger than the calculated one. Hence we can conclude that the hydrogen lines are formed not only in the hot spot, but rather in a region with a lower density. Thus, a theoretical reproduction of the observed spectrum of hydrogen lines is meaningless without taking into account radiation of circumstellar gas.
The parameters of the hot spot may be found without using hydrogen lines, in this case, to fit the absorption lines of the star and emission lines of He\,I -- He\,II to the observations, a relatively high density of the in-falling gas $N_0\sim10^{13}$ cm$^{-3}$ is needed (see the spectrum (2) on the Fig.\,\ref{TWHya} and \ref{TWHya1}), as it has  been already noted by Dodin et al. (2013), however at such densities the hydrogen lines are so broadened that their wings extend beyond the observed profiles.

The spectrum of hydrogen lines, arising in the hot spot, has not been calculated until now, therefore the conclusion that the hydrogen lines and the veiling degree in spectra of CTTS cannot be simultaneously reproduced with a single set of the accretion parameters $N_0,$ $V_0$ is new and important for diagnostics of AS.
It is quite natural from general considerations that in a real situation the accretion is not homogeneous.
Moreover, Dodin et al. (2013) as well as Ingleby et al. (2013) argue that observed spectra of CTTS are a sum of spectra, which are formed in different parts of the accretion region with various $N_0,$ $V_0.$
Using of the hydrogen spectrum for diagnostics makes us look at the problem from a new angle:
because the veiling continuum is undoubtedly formed in the hot spot and a high densities $\log N_0>12.0$ are required to explain the observed degree of the veiling, then there is a need to somehow weaken the hydrogen lines, which are formed in this region of (inhomogeneous) hot spot.

\section*{Conclusion}

The models of accretion hot spots on the surface of young stars are presented in this paper. The models are calculated taking into account departures from LTE for hydrogen and helium. At solar elemental abundances the next  in importance (after H and He) absorbers of the external ionizing radiation are the ions of C and O and their inclusion into the non-LTE treatment does not lead to significant changes in the structure of the hot spot.
Hence, it can be concluded that non-LTE-effects on the structure of the hot spot, which are related with differences in the absorption of the external radiation in the LTE and non-LTE case, are accounted for quite completely. Taking into account departures from LTE for other elements, which determine the opacity at optical wavelengths, can in principle lead to changes in the atmospheric model, but probably these changes should be the same as in the case of main sequence stars with effective temperatures about the temperature of the hot spot($<10000$~K), i.e. negligibly small (Hauschildt et al., 1999).

The Eddington flux $H_{\lambda}$ was calculated as a byproduct of the modelling of the atmospheric structure. The analysis of $H_{\lambda}$ allowed us to make a few new conclusions: the broad wings of hydrogen lines can be formed in the hot spot; the lines of "hot"{} ions such as C\,IV are almost not formed in the considered models.

We did not succeed to describe in a non-contradictory manner the observed spectrum of TW Hya using a single set of  $N_0,$ $V_0$ from a precalculated grid of spectra for a range of parameters $\log N_0=11.5-13,$ $ V_0=200-400$ km\,s$^{-1}$. This is due to the fact that to obtain the required ratio between the helium lines and continuum, we have to use models with $N_0\sim10^{13}$cm$^{-3}$, but in this case we obtain profiles of hydrogen lines, which are significantly broader than the observed ones. Perhaps this problem can be solved by assuming the simultaneous existence in the accretion zone the regions with $\log N_0<13$ and regions with a higher density $N_0,$ at which the in-falling gas becomes optically thick in the continuum. In this case the observer does not see the radiation of the heated atmosphere, because the photosphere of the shock is placed in the free-falling gas, i.e. in the pre-shock.

The spectrum of such gas must resemble the spectrum of the star with $T_{eff}\sim(F_{acc}/\sigma)^{0.25}\gtrsim10000$~K. It means that there must be signs of hydrogen and helium lines, which are in absorption and shifted to the red side. Such features are observed in spectra of some CTTS, in particular, the absorption feature in the hydrogen lines with a maximum depth at the radial velocity about +300~km\,s$^{-1}$ can be found in the presented spectrum of TW Hya. To verify this explanation, a more detailed (than it has been done in this work) modelling of the structure and spectra of the pre-shock region should be carried out for an extended range of $N_0$.  
Such modelling is a problem of current interest, since the lines forming in the in-falling gas can be informative for the diagnostics of the accretion and are already beginning to be used for these purposes (Petrov et al., 2014).

The calculated grid of models and spectra of the hot spot with taking into account departures from LTE for hydrogen and helium is available at the website\\
\href{http://lnfm1.sai.msu.ru/~davnv/hotspot}{\url{lnfm1.sai.msu.ru/~davnv/hotspot}}

~\\
{ACKNOWLEDGMENTS.} The author wishes to thank V.P.\,Grinin, S.A.\,Lamzin and L.I.\,Mashonkina for useful discussions;
the anonymous referees for notes, which allowed to improve the paper.  The author acknowledges support from "Dynasty"{} foundation and from the Program for Support of Leading Scientific Schools (NSh-261.2014.2). Based on observations made with ESO Telescopes at the La Silla Paranal Observatory under programme ID 085.C-0764(A).

		%%%%%%%%%%%%%%%%%%%%%%%%%%%%%%%%%%%%%%%%

\end{document}